\documentclass[12pt]{article}

\def\bea{\begin{eqnarray}} 
\def\eea{\end{eqnarray}} 
\newcommand{\gsim}{\stackrel{>}{_\sim}}

\title{The simplest little Higgs}
 
\author{\\
        M. Schmaltz \ \ \\ \\ 
        \small \sl \ Physics Department, Boston University, 
                Boston, MA  02215\\ \\ 
       }

\begin{document} 
\baselineskip=17pt 
\pagestyle{plain}

\begin{titlepage} 

\maketitle 

\begin{picture}(0,0)(0,0) 
\put(308,280){BUHEP-04-09}
\end{picture}

\begin{abstract} 
\leftskip-.3in 
\rightskip-.3in 
\vskip.3in 

We show that the $SU(3)$ little Higgs model
has a region of parameter space in which electroweak
symmetry breaking is natural and in which corrections to
precision electroweak observables are sufficiently small.
The model is anomaly free, generates a Higgs mass near
150 GeV, and predicts new gauge bosons and fermions at 1 TeV.

\end{abstract} 
\thispagestyle{empty} 
\setcounter{page}{0} 
\end{titlepage} 

\section{Introduction} 

The Standard Model Higgs mass suffers from quadratically divergent
quantum corrections which destabilize electroweak symmetry breaking.
In the presence of these divergences it is unnatural for the
Higgs vacuum expectation value and the $W$ and $Z$ 
masses to be lower than the cutoff by more than a factor of $4 \pi$.
Turning the argument around: we know the $W$ and $Z$ masses to
be around 100 GeV, therefore any natural extension of the
Standard Model must contain new physics at or below $\sim$ 1 TeV
in order to cancel the divergences.
It is exciting that the scale of new physics is within reach
of the LHC and possibly also the Tevatron.

Possibilities for this new physics include supersymmetry, technicolor,
extra dimensions, or the Little Higgs \cite{ACG2,Arkani-Hamed:2002qx,
Arkani-Hamed:2002qy,Gregoire:2002ra,Low:2002ws,Kaplan:2003uc,Chang:2003un,
Skiba:2003yf,Chang:2003zn,Cheng:2003ju}. New physics
contributes corrections to precision electroweak observables via
tree level exchange of new heavy particles or loop effects.
Experiments are so precise that they push the (indirect)
lower bounds on new particle masses
up to several TeV \cite{PDG,lepII,EW1,EW2}. 
This presents a challenge to model building: in order to avoid fine tuning
new physics is required near 1 TeV, but particles with TeV scale masses
are already severely constrained experimentally.

Recently, Cheng and Low \cite{Cheng:2003ju}
pointed out that little Higgs models can
be constructed with a symmetry (T parity)
that forbids all tree level contributions from the new physics
to electroweak observables while still allowing
the loops necessary to cancel divergences. T parity is
analogous to R parity in supersymmetric models.

In this paper we point out that the $SU(3)$ little Higgs model
proposed by Kaplan and Schmaltz \cite{Kaplan:2003uc}
allows a natural
solution of the ``little hierarchy problem'' without
T parity. We find that the model has regions of parameter space
for which TeV scale particles only couple very weakly
to Standard Model fields in tree level interactions. This
allows them to hide from precision electroweak measurements
while still canceling the divergences to the Higgs mass.
We find that new fermion and gauge boson masses as low as
1 TeV are consistent with the data. 

In the original $SU(3)$ model \cite{Kaplan:2003uc} anomalies
are not canceled in the low energy theory,
thus requiring new structure at the cut-off.
Here we present a different -- anomaly free -- choice of fermion
representations \cite{Pisano:ee}, which requires no spectators,
and provides a better fit to precision electroweak
measurements.
This anomaly free version of the $SU(3)$ model has also recently
been UV extended to $\sim 50$ TeV with another little Higgs
theory \cite{KSS}.

In the following section we review the essential ingredients
of the $SU(3)$ model. The weak interactions of the Standard Model
are enlarged from $SU(2)\times U(1)$ to $SU(3)\times U(1)$ at the
TeV scale. The new $SU(3)$ gauge bosons automatically
cancel quadratic divergences to the Higgs mass from $W$ and $Z$
loops. Furthermore, in making the top Yukawa coupling $SU(3)$
invariant one introduces a new particle which
also cancels the quadratic divergence to the Higgs
mass from the top loop. We also determine the Higgs potential
and show that top quark loops generate a
potential for the Higgs which leads to dynamical electroweak
symmetry breaking. Obtaining the correct Higgs VEV and a
sufficiently large Higgs mass ($m_H>114$ GeV) requires inclusion
of a tree level potential term similar to the $\mu$ term
in supersymmetry.

In the third Section we discuss naturalness and precision
electroweak constraints. This $SU(3)$ model differs
from other little Higgs models in that the Higgs quartic 
coupling is generated from radiative corrections at low
energies. This has the advantage that it automatically avoids
quadratically divergent contributions to the Higgs
mass from Higgs loops. On the flip side, it implies a small
quartic coupling which in turn requires a small Higgs mass
parameter to obtain the correct VEV. A small Higgs mass
combined with naturalness arguments requires the new physics
which cuts off quadratic divergences to be very light.
Interestingly, precision data do allow the new states in
this model to be light. We find a region in parameter
space with new gauge bosons and fermions as light as 1 TeV
and only relatively mild tuning of the Higgs mass parameter
of order 10\% (to be compared with $\sim$ 2\% in minimal supergravity).
This mild tuning can be avoided by also implementing
collective symmetry breaking in the Higgs potential. In this way
one can generate a tree level contribution to the quartic potential
at the cost of additional structure (see e.g. the $SU(4)$
model \cite{Kaplan:2003uc} or a model by
Skiba and Terning \cite{Skiba:2003yf}).

We conclude the paper with a brief discussion of the phenomenology.

\section{The model}

The underlying prescription for constructing the model is very simple.
Enlarge the SM $SU(2)_w \times U(1)_Y$ gauge group to
$SU(3)_w \times U(1)_X$ in a minimal way. This entails
enlarging $SU(2)$ doublets of the SM to $SU(3)$ triplets,
adding $SU(3)$ gauge bosons, and writing $SU(3)$ invariant
interactions which reproduce all the SM couplings when restricted to
SM fields. 

Explicitly, a SM generation is embedded in
$(SU(3)_c,SU(3)_w)_{U(1)_X}$ representations
\bea
\Psi_Q=(3,3)_{\frac13}\quad & \Psi_L=(1,3)_{-\frac13}\nonumber\\
d^c=(\bar 3,1)_{\frac13}\quad & e^c=(1,1)_1\nonumber  \\
2\times u^c=(\bar 3,1)_{-\frac23}\quad & n^c=(1,1)_0 
\eea 
where the triplets $\Psi_Q$ and $\Psi_L$ contain the quark and lepton
doublets and $u^c, d^c, e^c, n^c$ are the (charge-conjugated) singlets.%
\footnote{This charge assignment for the fermions leaves the
$SU(3)_w$ and $U(1)_X$ gauge groups anomalous. The anomalies
can be canceled by a Wess-Zumino term. This term is a higher
dimensional operator but does not become strongly coupled
until the cutoff $\Lambda=4 \pi f$. In order to avoid having to
cancel anomalies with spectator fermions at the cutoff we
present a different -- anomaly free -- embedding in the
section on fermions.}  
There are two $u^c$ fields, one is the SM right-handed up-type
quark, the other obtains a large mass with the third components of the
triplet $\Psi_Q$. Similarly, the singlet $n^c$ is the Dirac partner 
of the third component of $\Psi_L$.

The symmetry breaking,
$SU(3)_w \times U(1)_X \rightarrow SU(2)_w \times U(1)_Y$, is achieved 
with aligned vacuum expectation values for two complex triplet scalar fields
\bea
\Phi_1,\Phi_2= (1,3)_{-\frac13}.
\eea

The gauge interactions of the model are uniquely determined by gauge
invariance. SM Yukawa couplings and masses for the
heavy exotic fermions arise from the couplings
\bea
\lambda^u_1 u_1^c \Phi_1^\dagger \Psi_Q + 
\lambda^u_2 u_2^c \Phi_2^\dagger \Psi_Q +
\lambda^d {d^c \Phi_1 \Phi_2 \Psi_Q\over\Lambda} \nonumber \\ +
\lambda^n n^c \Phi_1^\dagger \Psi_L +
\lambda^e {e^c \Phi_1 \Phi_2 \Psi_L\over\Lambda}
\label{eq:yukawas}
\eea
after decomposing the fields into $SU(2)_w\times U(1)_Y$ multiplets.
Note that the $SU(3)_w$ indices of the rightmost terms are contracted with
epsilon tensors. 
We will discuss the detailed structure of these couplings when we need
them for the potential computations in the next section. 

We find it convenient to work with non-linear sigma model fields
$\Phi_i$ which can be obtained
from normal complex triplets with vacuum expectation values $f_1$
and $f_2$ by integrating out the radial modes. The non-linear
sigma model is more general as it may arise from many different
UV completions, the linear sigma model is only one example.
5 of the 10 degrees of freedom in the $\Phi_i$ are eaten by
the Higgs mechanism when $SU(3)_w$ is broken.
We parameterize the remaining degrees of freedom as  
\bea
\Phi_1= e^{i\Theta  {f_2\over f_1} }
\left( \begin{array}{l}
0  \\ 0 \\ f_1 \end{array} \right) , \quad
\Phi_2= e^{-i \Theta {f_1\over f_2}}
\left( \begin{array}{l}
0  \\ 0 \\ f_2\end{array} \right) 
\label{eq:phis}
\eea
where
\bea
\Theta = \frac1{f}\left[{\eta\over \sqrt{2}} + 
\left( \begin{array}{cc} 
\!\!\begin{array}{ll} 0 & 0 \\ 0 & 0 \end{array} 
& \!\!h \\ h^\dagger  & \!\!0 \end{array} \right)\right]
 \quad  {\rm and}\quad  f^2 = f_1^2+f_2^2 \ .
\label{eq:phiexpand}
\eea
Here the field $h$ is an $SU(2)_w$ doublet which we identify
with the SM Higgs doublet and $\eta$ is a real scalar field.
Their normalizations are chosen to produce canonical kinetic terms. 

\subsection{Gauge bosons}

In addition to the SM gauge bosons our model contains 5 new
gauge bosons with masses of order the scale $f$. The
new gauge bosons fill out a complex $SU(2)\times U(1)$ doublet
$(W'_+,W'_0)$ with
hypercharge $\frac12$ and a neutral singlet $Z'$.

The gauge boson masses are determined from the kinetic terms of the $\Phi_i$
\bea
|(\partial_\mu+i g A^a_\mu T^a -\frac{i}3 g_x A_\mu^x)\Phi_i|^2
\rightarrow
tr [(g A^a_\mu T^a -\frac{1}3 g_x A_\mu^x)^2\Phi_i \Phi_i^\dagger]
\eea
To compute the masses it is convenient to use the $3\times 3$ matrix
\bea
<  \Phi_1 \Phi_1^\dagger + \Phi_2 \Phi_2^\dagger > =
\left( \begin{array}{cc}
\!<\!\!h h^\dagger\!\!> & 0  \\ 
0 & f^2 
\end{array}\right)
\eea
expanded out to leading non-trivial order in $h/f$, and tracing it
with the squares of the generators $T^a$ and $T^x$. In our conventions
$<\!h^T\!>=(\frac{v}{\sqrt{2}}\, 0)$.
To leading order the masses for the $W_\pm$ and the $SU(2)$ doublet of 
heavy gauge bosons $(W'_\pm, W'_{0,\overline 0})$ are 
\bea
m^2_{W_\pm}&=&\frac{g^2}4 v^2 \nonumber \\
m^2_{W'_\pm}&=&\frac{g^2}2 f^2 \nonumber \\
m^2_{W'_0}&=&\frac{g^2}2 f^2 \ .
\eea

The mass matrix for the neutral gauge bosons is more
complicated. After $SU(3)\times U(1)_X$ breaking the neutral
gauge boson corresponding to the diagonal
$SU(3)$ generator $T^8=\frac1{\sqrt{3}}\, diag(\frac12,\frac12,-1)$ and the
$U(1)$ generator $T^x$ mix.
The mass eigenstates (before taking into account the Higgs VEV) are
\begin{eqnarray}
W^3_\mu &=& A^3_\mu 
\nonumber\\
B_{\mu} &=& {-g_x A_{\mu}^{8} + \sqrt{3}g B_{\mu}^x \over
\sqrt{3g^2 + g_x^2}} \nonumber \\
Z'_{\mu} &=& {\sqrt{3}g A_{\mu}^{8} + g_x B_{\mu}^x \over
\sqrt{3g^2 + g_x^2}} 
\end{eqnarray}
and the hypercharge gauge coupling is
\bea
g'=g_x/ \sqrt{1+\frac{g_x^2}{3g^2}} \ . 
\eea

After $SU(2)\times U(1)_Y$ breaking the photon remains massless as
in the SM, but the $Z$ mixes with the heavy $Z'$ which leads to
small deviations from the SM which we discuss in Section 3.
The resulting masses are
\bea
m_Z^2&\!\!=\!\!& \frac{g^2}4 v^2 (1+t^2)
\nonumber \\
m_{Z'}^2&\!\!\!=\!\!\!& {g^2}f^2 {2\over 3\!-\!t^2} 
\eea
where $t=g'/g=\tan{\theta_W}$ and $\theta_W$ is the weak mixing angle.

\subsection{Fermions}

In this section we describe two different embeddings of the quarks and
leptons in $SU(3)_w \times U(1)_X$.

{\it Model 1:} All three generations carry identical gauge quantum
numbers and the $SU(3)_w \times U(1)_X$ gauge group is anomalous.
Fermion masses for all generations arise from the Yukawa couplings of
equation (\ref{eq:yukawas}). The SM down type Yukawa matrix
is equal to the matrix $\lambda_d\, {f/ \Lambda}$
which can be seen easily by expanding out
$\epsilon\, \Phi_1\Phi_2/\Lambda \rightarrow
\left((\sigma_2 h)^T \ 0\right)\, f/\Lambda$.

The up-type Yukawa matrices are more interesting as there
are 6 quarks of charge 2/3 which mix with each other.
In general this leads to flavor changing neutral currents and
is dangerous. We assume that one of the matrices
$\lambda_i^u$ is approximately proportional to the unit matrix.
This assumption makes the theory completely safe from flavor
changing effects and can probably be relaxed somewhat.

We choose the matrix $\lambda^u_2$ to be proportional to ${\bf 1}$.
$\lambda_1^u$ must then be hierarchical in order to produce
hierarchical SM quark masses,
it can be diagonalized by unitary transformations
on the fields $\Psi_Q$ and $u_1^c$. In this basis the $6\times 6$
mass matrix for charge 2/3 quarks decouples into three $2\times 2$
matrices, each describing the mixing of a SM up-type quark
$(u,c,t)$ and it's heavy partner $(U,C,T)$.
Explicitly, the mass term is
\bea
\left( \, u_1^c \  u^c_2 \right)
\left( \begin{array}{c} 
\lambda^u_1 \Phi_1^\dagger \\ 
--- \\
\lambda^u_2 \Phi_2^\dagger \end{array} \right)
\left( \! \Psi_{Q} \begin{array}{c} \! \\ \! \end{array}\!\!\!\!\! \right)
\label{eq:fermass}
\eea
where we suppressed generation indices.
After substituting VEVs for the $\Phi_i$
the $2\times 2$ submatrices for each generation are
\bea
\left( \begin{array}{rr}
\lambda^u_1 \!<\!\!h\!\!>\!\!f_2/f &
\lambda^u_1 f_1 \\ 
-\lambda^u_2 \!<\!\!h\!\!>\!\!f_1/f& 
\lambda^u_2 f_2 \end{array} \right)
\eea
Diagonalizing, we find the masses of the up-type quarks and their partners
\bea
m_u&=&\lambda_u <\!\!h\!\!>
\nonumber \\
m_U&=& \sqrt{(\lambda^u_1 f_1)^2 + (\lambda^u_2 f_2)^2 }
\label{eq:topyukawa}
\eea
where we have defined 
\bea
\lambda_u&=&\lambda^u_1 \lambda^u_2 \frac{f}{m_{U}} \ .
\eea
For the first two generations
$\lambda^u_1 \ll \lambda^u_2$, and these expressions
further simplify to $m_u=\lambda^u_1 <\!\!h\!\!>$
and $m_U=\lambda^u_2 f_2$.
Diagonalizing the mass matrices mixes the 
up-type quarks fields in $\Psi_Q$ with their
$SU(2)$ singlet partners
by an amount 
\bea
\theta_u=<\!\!h\!\!>\!\!f_1/(f_2 f)
\label{eq:fermionmixing}
\eea
which leads to shifts of order $\theta_u^2$ in
the $W$ and $Z$ couplings of the up-type quarks and neutrinos.

{\it Model 2:}
In the second model we cancel the $SU(3)_w$ anomaly by taking
different charge assignments for the different generations
of quark triplets
\bea
\Psi_{Q^3}=(3,3)_{\frac13}\quad &
\Psi_{Q^{1,2}}=(3,\bar 3)_{0}\quad &
\Psi_L=(1,3)_{-\frac13}\nonumber \\
d^{c 3}=(\bar 3,1)_{\frac13}\quad &
2\times d^{c 1,2}=(\bar 3,1)_{\frac13}\quad &
e^c=(1,1)_1\nonumber  \\
2\times u^{c 3}=(\bar 3,1)_{-\frac23}\quad &
u^{c 1,2}=(\bar 3,1)_{-\frac23}\quad
& n^c=(1,1)_0 
\eea 
where the superscripts label generations. The leptons are unchanged
from the previous model. With this new charge assignment all
anomalies cancel \cite{Pisano:ee}
which makes this model easier to UV complete \cite{KSS}.
Note that now the heavy quarks ($T, S, D$) are partners of
the top, strange and down quarks, respectively. 
The quark Yukawa couplings stem from operators of the form
\bea
u^{c\, i} \Phi^\dagger \Psi_{Q^3} + 
{u^{c\, i} \Phi^\dagger \Phi^\dagger \Psi_{Q^{1,2}}\over\Lambda} +
d^{c\, i} \Phi \Psi_{Q^{1,2}} +
{d^{c\, i} \Phi \Phi \Psi_{Q^3}\over\Lambda} +
\label{eq:newyukawas}
\eea
where we have suppressed indices labeling the $\Phi_i$ and
the two copies of conjugate fields. As in the previous model these
operators allow general $3\times 3$ Yukawa 
matrices for up and down quarks and leptons.
The diagonalization of mass matrices is different from the
previous model as there is only one new up-type quark and two
new down-type quarks and associated mixing angles $\theta_d$ with the
light quarks. 
In order to avoid flavor changing neutral currents the mass
matrix of heavy partners needs to be sufficiently
well aligned with the quark mass matrices to avoid
flavor changing neutral currents. We expect the
constraints from flavor physics to be interesting
and non-trivial \cite{Sekhar} but a detailed
study is beyond the scope of this paper.

Note that in both models the mixing of light fermions with their
partners generates a coupling of the $W$ and $Z$ gauge bosons 
to a single SM fermion and it's heavy partner proportional
to $\theta_{u,d}$. This opens the interesting possibility
of single $U,D$ production from fusion of weak gauge bosons
with SM quarks (e.g. $d+W\rightarrow U,\  u+Z\rightarrow U$).

\subsection{Scalars}

The two scalar triplets $\Phi_i$ which are responsible
for $SU(3)\times U(1) \rightarrow SU(2)\times U(1)$
breaking contain 10 real degrees of freedom. 5 are
eaten by the $SU(3)$ gauge bosons with TeV scale
masses, 4 form the SM Higgs doublet $h$ and one
is a real scalar field $\eta$.
Since we did not include an operator which gives a quartic
coupling for the Higgs, this must be generated dynamically.
Explicitly, the radiative corrections should produce
the standard model Higgs potential
\bea
V=m^2 h^\dagger h + \lambda (h^\dagger h)^2
\label{eq:higgspot}
\eea 

Electroweak symmetry breaking requires $m^2$ negative and
of order of the electroweak scale, and the LEP bound on the
Higgs mass requires $\lambda \gsim 0.11$.
We now discuss the form of the radiative contributions to the
potential.
Above the scale $f$
the $SU(3)$ gauge symmetry is unbroken and the potential
is best described in terms of the $SU(3)$ 
multiplets $\Phi_i$, and it is easy to see that the most general
potential is a function of the only gauge invariant which depends on
the Higgs, $\Phi_1^\dagger \Phi_2$.
At the scale $f$, the $SU(3)$ partners of fermions
and gauge bosons obtain masses, and the theory matches onto the
standard model. Below $f$ the Higgs potential
receives the usual radiative corrections
from top quark and gauge loops.

We first discuss the potential generated above the scale $f$. The 
top Yukawa couplings and gauge couplings preserve a
$U(1)$ symmetry under which $\Phi_1$ and $\Phi_2$ have opposite charges.
Therefore the lowest dimensional operator which
can be radiatively induced is $|\Phi_1^\dagger \Phi_2|^2$.
This operator is already
generated at one loop but only with a logarithmic divergence.
Its contribution
to the Higgs mass is of order $f^2/(16 \pi^2) \sim m_W^2$.
The symmetry forbids any quadratically divergent contributions
from gauge or Yukawa couplings.

As can be seen from the explicit formulae below, the radiatively
generated potential alone generates a Higgs ``soft mass squared''
which is somewhat too large. Therefore we also include
a tree level ``$\mu$" term which will partially cancel the
Higgs mass. It explicitly breaks
the spontaneously broken global $U(1)$ symmetry and
gives a mass to the would-be Nambu-Goldstone boson $\eta$
\bea
V_{tree}=\mu^2 \Phi_1^\dagger \Phi_2 + h.c. \rightarrow
\mu^2 {f^2\over f_1 f_2} (h^\dagger h+\frac12 \eta^2) 
-\frac1{12} {\mu^2 f^4\over f_1^3 f_2^3} (h^\dagger h)^2 + \dots
\label{eq:muterm}
\eea
Since the operator contains a Higgs mass, $\mu$ must be near the
weak scale $m_W$. The mass scale $\mu$ is radiatively
stable because this is the only $U(1)$ breaking operator in the theory.
 
We compute the radiatively generated one-loop potential from gauge and
Yukawa interactions using the
formalism of Coleman and Weinberg \cite{CW}. Details of the calculation
are given in an Appendix. The corrections to the potential Eq.~(\ref{eq:higgspot})
contain a mass squared $\delta m^2$ and a quartic $\delta \lambda$

\bea
\delta m^2\!=\!\frac{-3}{8\pi^2}\!\!\left[\!\lambda_t^2 m_{T}^2
Log\!\left(\!\frac{\Lambda^2}{m_{T}^2}\!\right)\!
\!-\!\frac{g^2}{4} m_{W'}^2
Log\!\left(\!\frac{\Lambda^2}{m_{W'}^2}\!\right)\!
\!-\!\frac{g^2}{8} (1\!+\!t^2) m_{Z'}^2
Log\!\left(\!\frac{\Lambda^2}{m_{Z'}^2}\!\right)\! \right]\nonumber
\eea
\bea
\delta \lambda = {|\,\delta m^2| \over 3} \frac{f^2}{f_1^2 f_2^2} \nonumber \\
+\frac{3}{16\pi^2}\!\!&\!\!&\!\!\hskip-.2in\left[
\lambda_t^4 Log\!\left(\frac{m_{T}^2}{m_t^2}\right)
-\frac{g^4}{8} Log\!\left(\!\frac{m_{W'}^2}{m_W^2}\!\right)
-\frac{g^4}{16} (1\!+\!t^2)^2 Log\!\left(\frac{m_{Z'}^2}{m_Z^2}\right)
\! \right]  \nonumber 
\eea

Assuming that there are no large direct contributions to the potential
from physics at the cutoff we have
\bea
V_{total}=(\mu^2 {f^2\over f_1 f_2} + \delta m^2) h^\dagger h +
( -\frac1{12} {\mu^2 f^4 \over f_1^3 f_2^3} + \delta \lambda) (h^\dagger h)^2
\label{eq:higgspotmu}
\eea
Note that the radiative contribution to the Higgs mass from the top loop
is negative while the contribution to the quartic is positive. Thus we have
radiative electroweak symmetry breaking and stability of the Higgs potential.

The potential depends on a number of free parameters, before analyzing it
further we first determine what ranges for the parameters are reasonable
by studying the constraints from precision electroweak data.

\section{Electroweak constraints and naturalness}

The model predicts small changes to the masses and
couplings of the Z from Z-Z' mixing and to the fermion
couplings from q-Q mixing. Both types of corrections
scale as $v^2/f^2$ and decouple as we take $f$ large.
Naturalness arguments prefer $f$ to be as near to the TeV
scale as possible, therefore a detailed a study of the size of the
corrections is needed. This study has been performed for a model
\cite{Skiba:2003yf} which closely related to our model 1. We quote
some of the results and add constraints from LEP II. 

To start, note that the numerically most significant contribution
to the Higgs mass comes from the top loop, therefore naturalness
arguments primarily lead to an upper bound on the mass of the heavy
partner of the top quark
\bea 
m_T=\sqrt{(\lambda_1^t f_1)^2 + (\lambda_2^t f_2)^2 } \ .
\eea
On the other hand, deviations from the standard model in
precision electroweak physics stem mostly from mixing with
the Z' and from four fermion operators mediated by the Z'.
Thus precision electroweak physics implies a lower bound on
the mass of the Z'
\bea
m_{Z'}^2={2 g^2 \over 3-t^2} (f_1^2 + f_2^2)
\eea
We see that it is possible to lower the mass of the top partner
while keeping the Z' mass fixed by going to a region in parameter
space in which the $f_i$ are different. Then the Z' mass is dominated
by the larger of the two $f_i$ whereas $m_T$ can be made smaller
by reducing the corresponding Yukawa coupling $\lambda^t_i$. 

To reduce the number of parameters we determine the $\lambda_i^t$
such that the $T$-mass is minimized for given scales $f_i$
and top Yukawa coupling Eq.~(\ref{eq:topyukawa}).
This gives the values
\bea 
\lambda_1=\sqrt{2} \lambda_{top} {f_2 \over f} \quad\quad
\lambda_2=\sqrt{2} \lambda_{top} {f_1 \over f} \quad\quad
m_T=2 \lambda_{top} {f_1 f_2 \over f} 
\eea
The $T$-mass is not very sensitive to this precise choice, i.e. it is
not fine-tuned.

In the following, we imagine the scales $f_i$ to differ by a factor
of a few with $f_2 > f_1$. This choice reduces
fermion mixing Eq.~(\ref{eq:fermionmixing}) sufficiently, such
that no significant deviations from the standard model arise, and
we ignore fermion mixing for the
precision electroweak analysis.

The largest deviations from the standard model derive from
tree level Z-Z' mixing and Z' induced four fermion operators.
We find the custodial $SU(2)$ symmetry violating shift in
the Z mass 
\bea
\delta\rho \equiv \alpha T= \frac18 \frac{v^2}{f^2} (1-t^2)^2\ ,
\eea
the modified Z-couplings
\bea
{e \over c_w s_w} Z^\mu \left[
J^3_\mu - s_w^2 J^Q_\mu - {v^2 \over 8 f^2}
(1-t^2)(\sqrt{3}J^8_\mu + t^2 J^Y_\mu) 
\right] \ ,
\eea
and the Z' induced four fermion operators
\bea
\delta {\cal L}= - { (\sqrt{3}J^8_\mu + t^2 J^Y_\mu)^2 \over 4 f^2} \ .
\label{eq:fourfermi}
\eea
Here $s_w=sin\,\theta_w$, $c_w=cos\,\theta_w$, and the fermion
currents $J_\mu$ are defined such that the neutral gauge bosons couple
as $g A_3 J^3 + g A_8 J^8 + g_x A_x J^x$, 
and $J^Y\equiv J^x-J^8/\sqrt{3}$, $J^Q\equiv J^3+J^Y$.
Note that these formulae apply to both fermion embeddings.
As is customary we use the experimentally measured values
for $\alpha_{em}, s_w, G_F$ to fix the parameters $g,s_w,v$ in
our lagrangian. 

We begin by looking at the bound implied by the Z-mass shift
which corresponds to a non-vanishing T parameter. Applying
the 95\% confidence limit $T<0.15$ (for a light Higgs) \cite{PDG}
we obtain $f\gsim 2.0$ TeV in either model.

LEP II data constrain the possible contribution to $e^+ e^-$ scattering
from four fermion operators \cite{lepII}.
We find that the best constraint comes
from considering the operator $(\overline e \gamma^\mu \gamma^5 e)^2$
contained in Eq.~(\ref{eq:fourfermi})
which gives $f\gsim 2.0$ TeV at the 95\% confidence level in both models.

The strongest constraint in the analysis of \cite{Skiba:2003yf}
is due to new contributions to atomic
parity violation, both from the shift in the Z coupling
as well as the four Fermi operators. Since the couplings of
the first generation quarks to  the Z and Z' differ in the two
models we find different limits. In model 1
the magnitude of the predicted ``weak charge'' of Cesium is increased
proportional to $v^2\over f^2$. Using the experimental value
\cite{PDG,Wood:zq} $Q_W^{exp}=-72.69\pm .48$ and the Standard
Model prediction
$Q_W^{SM}=-73.19\pm.03$ we find a strong limit $f\gsim 3.9$ TeV
at 95\% confidence level. 
In model 2 the weak charge is reduced due to the new interactions.
This improves the fit and leads to a bound $f\gsim 1.7$ TeV
(at 95\% confidence).

The results for model 1 are similar to the results of \cite{Skiba:2003yf}
who performed a fit to all precision electroweak
measurements and find $f>3.3$ TeV at 95\% confidence,
and $f\gsim 2.5$ TeV if the
Cesium APV constraint (which depends on difficult to quantify
systematic errors in atomic physics calculations) is dropped.

We summarize that there is significant tension between precision
electroweak constraints and naturalness in model 1. Model 2
avoids all constraints in the region of parameter space with
$f\sim f_2 \gsim 2$ TeV and $f_1$ somewhat smaller. For example,
picking the ``golden'' point, $f_1=0.5$ TeV and $f_2=2$ TeV, we find
\bea
m_T=1.0\, {\rm TeV} \quad \quad
m_D=m_S=0.7\, {\rm TeV} \nonumber \\
m_{Z'}=1.1\, {\rm TeV} \quad \quad \quad \quad
m_{W'}=0.9\, {\rm TeV}
\eea

For this point we can now also determine the approximate Higgs mass
from Eq.~(\ref{eq:higgspotmu}). Fixing the Higgs VEV at 246 GeV,
the cutoff $\Lambda = 5$ TeV ($\sim 4 \pi f_1$), we find 
$m_{Higgs}=140$ GeV. The amount of fine tuning can be defined using
the sensitivity of the squared Higgs VEV $v^2$ to the parameter
$\mu^2$,
i.e. ${\mu^2 \over v^2}{\partial v^2 \over \partial \mu^2}$.
We find a sensitivity of about 10 for the parameter choice
above which corresponds to 10\% tuning.

\section{Conclusions}

We have shown that the $SU(3)$ simplest little Higgs model 
has a natural region of parameter space in which precision
electroweak constraints are satisfied and
electroweak symmetry breaking only requires mild tuning
of order 10\% (to be compared with $\sim 2\%$ for the MSSM).

It would be interesting to explore flavor changing effects
mediated by the new particles. We expect nontrivial constraints
because the couplings of new physics to the third generation
differs from couplings to the first two generations.

LHC phenomenology for this model is exciting as it promises
new gauge bosons and fermions near 1 TeV, both within reach of
the LHC \cite{pheno1,pheno2}.
Easiest to produce and detect is the Z' which
can be singly produced and has a significant
branching ratio into lepton pairs. Heavy quarks can be produced
singly via fusion of a $u$ or $d$ quark with a weak gauge boson
(in analogy with single top production) or in pairs from
strong interactions.

\section{Acknowledgments}

I thank Kaladi S. Babu for collaboration and inspiration at
the beginning of this project. I also thank David E. Kaplan
and Witold Skiba
for numerous helpful discussions and their hospitality
at Yale and The Johns Hopkins University.
MS acknowledges support from an Alfred P. Sloan Research
Fellowship and a DOE Outstanding Junior Investigator award.

\appendix

\section{Appendix: Coleman-Weinberg potential}

We compute the one-loop radiatively generated Higgs potential
using the method of Coleman and Weinberg \cite{CW}.
One substitutes the Higgs by its vacuum expectation value
and computes the vacuum energy at one loop as a function of the
particle masses. Since the particle masses depend on the Higgs
expectation value, this determines the Higgs dependence of the
vacuum energy: the Higgs potential.
At one loop we only need to compute loops with no interactions,
summed over all the fields in the theory. 

\subsection{Fermion contribution}

The only numerically relevant contribution from fermion loops
is due to the top quark and its partner. To compute it we need
to determine the top quark mass matrix in the presence of
$SU(2)_w$ breaking. The top Yukawa coupling comes from
\bea
{\cal L}_{top} = 
(\lambda_1 t_1^c \Phi_1^\dagger + \lambda_2 t_2^c \Phi_2^\dagger)\  \Psi_T\ .
\eea
For the CW potential we need the hermitian mass squared matrix
\bea
M_{t T}^2=
\left( \begin{array}{c} 
\lambda_1 \Phi_1^\dagger \\ 
---- \\
\lambda_2 \Phi_2^\dagger \end{array} \right)
\left( \begin{array}{ccc} &\!\!|\!\!& \\
\!\!\lambda_1 \Phi_1\!\!&\!\!|\!\!&\!\!\lambda_2 \Phi_2\!\! \\&\!\!|\!\!&\end{array} \right)=
\left( \begin{array}{ll}
\lambda_1^2 \Phi_1^\dagger \Phi_1 &
\lambda_1 \lambda_2 \Phi_1^\dagger \Phi_2 \\ 
\lambda_2 \lambda_1 \Phi_2^\dagger \Phi_1& 
\lambda_2^2 \Phi_2^\dagger \Phi_2 \end{array} \right)
\eea
Expanding the $\Phi_i$ by using Eq.~[\ref{eq:phis}] and diagonalizing
we find the masses squared of the $t$ and $T$ quarks to fourth
order in the Higgs VEV. We use
the notation $m^2_t$ for the leading order (in $v^2/f^2$) top mass
and $m^2_{t,4}$ for the $4$-th order expression.
\bea
m_{t,4}^2&=&\lambda_t^2 <h^\dagger h> -
\left[ \frac13\, \lambda_t^2 \frac{f^2}{f_1^2 f_2^2} -
\frac{\lambda_t^4}{m_{T}^2}\right] <h^\dagger h>^2
\nonumber \\
m_{T,4}^2&=&m_{T}^2 - m_{t,4}^2
\eea
where we have used
\bea
m_{T}^2&=&\lambda_1^2 f_1^2 + \lambda_2^2 f_2^2 \nonumber \\
\lambda_t&=&\lambda_1 \lambda_2 \frac{f}{m_{T}}
\eea
We use the Coleman-Weinberg formula for fermions
with mass matrix $M_f$
\bea
V_{fermion}=-\frac{N_c}{16\pi^2} \Lambda^2 tr [ M_{f}^2 ] +
\frac{N_c}{16\pi^2} tr [M_{f}^4 Log\!\left(\frac{\Lambda^2}{M_{f}^2}\right)]
\eea
and see that the quadratic divergence cancels.
The Log-divergent piece gives
\bea
V_2&=&-\frac{3}{8\pi^2}\lambda_t^2 m_{T}^2
Log\!\left(\frac{\Lambda^2}{m_{T}^2}\right)\ h^\dagger h \nonumber \\
V_4&=&\frac{3}{16\pi^2}\left[
\lambda_t^4 Log\!\left(\frac{m_{T}^2}{m_t^2}\right)+
\frac23\frac{f^2}{f_1^2 f_2^2} \lambda_t^2 m_{T}^2
Log\!\left(\frac{\Lambda^2}{m_{T}^2}\right) \right]\ (h^\dagger h)^2
\eea

The form of the Logs is suggestive
toward an effective field theory interpretation. The
$Log\, (\Lambda^2/m_{T}^2)$ arises from
renormalization above the $T$ mass where
the theory is $SU(3)$ symmetric. Therefore an $SU(3)$ symmetric
potential is generated
\bea
V_{SU(3)}\!\!&\!\!=\!\!&\!
\frac{3}{8\pi^2}\, \lambda_1^2 \lambda_2^2 |\Phi_1^\dagger \Phi_2|^2 
Log\!\left(\frac{\Lambda^2}{\mu^2}\right) + {const.}\\
&\!\!=\!\!&\!\frac{3}{8\pi^2}\! \left[ - \lambda_t^2 m_T^2\, h^\dagger h +
\frac13 \lambda_t^2 {m_T^2 f^2 \over f_1^2 f_2^2} (h^\dagger h)^2 + \dots
\right]\!Log\!\left(\frac{\Lambda^2}{m_T^2}\right)+ {const.} \nonumber
\eea
where in the last line we replaced $\mu^2\rightarrow m_T^2$.
Below $m_T$, only the top quark remains in the
effective theory and produces its usual contribution to the 
quartic coupling proportional to
$Log\, ({m_{T}^2/m_t^2})$.

\subsection{Gauge boson contribution}

The gauge boson masses stem from the kinetic terms
of the $\Phi_i$
\bea
|(\partial_\mu+i g A^a_\mu T^a -\frac{i}3 g_x A_\mu^x)\Phi_i|^2
\rightarrow
tr [(g A^a_\mu T^a -\frac{1}3 g_x A_\mu^x)^2\Phi_i \Phi_i^\dagger]
\eea
We begin by computing the matrix
\bea
\Phi_1 \Phi_1^\dagger\!\!&\!\!+\!\!&\!\!\Phi_2 \Phi_2^\dagger = \nonumber \\
&&\hskip-.4in \left( \begin{array}{c}
h h^\dagger + (h h^\dagger)^2
(1/f^2-f^2/(3 f_1^2 f_2^2)) \hskip1.6in 0  \\ 
0 \hskip1.2in f^2 - h^\dagger h - (h^\dagger h)^2
(1/f^2-f^2/(3 f_1^2 f_2^2))  
\end{array}\right)
\eea
to the relevant order. We then trace this matrix with the squares of
the gauge generators and find the masses for the
$W$'s and an $SU(2)$ doublet of 
heavy gauge bosons $(W_{\pm}', W_{0,\overline 0}')$
\bea
m^2_{4,W_\pm}&=&\frac{g^2}4 v^2
[1 +{v^2\over 2 f^2}(1-\frac13 {f^4\over f_1^2 f_2^2})]\nonumber \\
m^2_{4,W_{\pm}'}&=&\frac{g^2}2 f^2 - m^2_{4,W_\pm} \nonumber \\
m^2_{4,W_{0}'}&=&\frac{g^2}2 f^2
\eea
The neutral gauge boson masses are more
complicated because the gauge bosons corresponding to the
$SU(3)$ generator $T^8\!=\!\frac1{2\sqrt{3}}\, diag (1,1,-2)$ and the
$U(1)$ generator $T^x$ mix.
The mass eigenstates before $SU(2)$ breaking are
\begin{eqnarray}
W^3_\mu &=& A^3_\mu 
\nonumber\\
B_{\mu} &=& {-g_x A_{\mu}^{8} + \sqrt{3}g B_{\mu}^x \over
\sqrt{3g^2 + g_x^2}} \nonumber \\
Z'_{\mu} &=& {\sqrt{3}g A_{\mu}^{8} + g_x B_{\mu}^x \over
\sqrt{3g^2 + g_x^2}} 
\end{eqnarray}
and the hypercharge gauge coupling is
$g'=g_x/ \sqrt{1+{g_x^2/3g^2}}\,$ .

After $SU(2)\times U(1)_Y$ breaking the photon remains massless as
in the SM, however the $Z$ mixes with the $Z'$ which leads to
deviations from the SM.
Using $t=g'/g=\tan{\theta_W}$ where $\theta_W$ is the weak mixing
angle we write the masses for the $Z$ and $Z'$ as
\bea
m_{4,Z}^2&\!\!=\!\!& \frac{g^2}4 v^2 (1+t^2)
[1 +{v^2\over 2 f^2}
(1-\frac13 {f^4\over f_1^2 f_2^2})- \frac{v^2}{8f^2} (1-t^2)^2]
\nonumber \\
m_{4,Z'}^2&\!\!\!=\!\!\!& \frac{g^2}{2}f^2{4\over 3\!-\!t^2} - m_{4,Z}^2
\eea
We insert these masses into the Coleman-Weinberg formula 
\bea
V_{gauge}=\frac{3}{32\pi^2} \Lambda^2 tr [ M_{g}^2 ] -
\frac{3}{64\pi^2} tr [M_{g}^4 Log\!\left(\frac{\Lambda^2}{M_{g}^2}\right)]
\eea
to obtain the contribution to the Higgs potential.
As in the fermion case the Higgs dependence in the quadratically
divergent term cancels. Summing over $W_\pm$, $W_\pm^{'}$, and $W_0^{'}$
the Log-divergent term gives
\bea
V_2&=&\frac{3}{32\pi^2} g^2 m_{W'}^2
Log\!\left(\frac{\Lambda^2}{m_{W'}^2}\right)\ h^\dagger h \\
V_4&=&-\frac{3}{128\pi^2}g^4\left[
Log\!\left(\frac{m_{W'}^2}{m_W^2}\right)+
\frac23\frac{f^4}{f_1^2 f_2^2}
Log\!\left(\frac{\Lambda^2}{m_{W'}^2}\right) \right]\ (h^\dagger h)^2
\nonumber
\eea
and from the $Z$ and $Z'$ we find
\bea
V_2\!\!&\!\!=\!\!&\!\!\frac{3}{64\pi^2} g^2 (1+t^2) m_{Z'}^2
Log\!\left(\frac{\Lambda^2}{m_{Z'}^2}\right)\, h^\dagger h \\
V_4\!\!&\!\!=\!\!&\!\!-\frac{3}{256\pi^2}g^4\left[(1+t^2)^2
Log\!\left(\frac{m_{Z'}^2}{m_Z^2}\right)+ \frac83
{1+t^2\over 3-t^2} \frac{f^4}{f_1^2 f_2^2}
Log\!\left(\frac{\Lambda^2}{m_{Z'}^2}\right) \right](h^\dagger h)^2
\nonumber
\eea

\noindent
Finally, summing contributions from both top and gauge sectors we have
\bea
\delta V=\delta m^2 h^\dagger h + \delta\lambda (h^\dagger h)^2
\eea
where 
\bea
\delta m^2\!=\!\frac{-3}{8\pi^2}\!\left[\!\lambda_t^2 m_{T}^2
Log\!\left(\!\frac{\Lambda^2}{m_{T}^2}\!\right)\!
\!-\!\frac{g^2}{4} m_{W'}^2
Log\!\left(\!\frac{\Lambda^2}{m_{W'}^2}\!\!\right)\!
\!-\!\frac{g^2}{8} (\!1\!+\!t^2) m_{Z'}^2
Log\!\left(\!\frac{\Lambda^2}{m_{Z'}^2}\!\right)\! \right]\nonumber
\eea
and
\bea
&\!\!\!\!\delta \lambda&\!\!\!\! = -{m^2 \over 3} \frac{f^2}{f_1^2 f_2^2}
+\frac{3}{16\pi^2} \times \nonumber \\
&&\!\!\!\!\!\!\!\!\!\!\!\!\left[
\lambda_t^4 (Log\!\left(\!\frac{m_{T}^2}{m_t^2}\!\right)\!-\!\frac12)
-\frac{g^4}{8} (Log\!\left(\!\frac{m_{W'}^2}{m_W^2}\!\right)\!-\!\frac12)
-\frac{g^4}{16} (1\!+\!t^2)^2
(Log\!\left(\!\frac{m_{Z'}^2}{m_Z^2}\!\right)\!-\!\frac12)
\right]  \nonumber 
\eea
where in the last line we have also restored the finite terms
which one obtains from expanding out the masses in the logarithms
in the Coleman-Weinberg formula in terms of the Higgs field.
In addition there are
finite cut-off dependent terms in the CW potential
$tr M^4 [Log(\Lambda^2/M^2) + 3/2]$, they can easily be included
by redefining $\Lambda \rightarrow 2.1 \, \Lambda$ in
these formulas.

\end{document}